\documentclass[a4paper,12pt]{article}
\usepackage[cp1251]{inputenc}
\usepackage[russian]{babel}

\newcommand{\be}{\begin{equation}}
\newcommand{\ee}{\end{equation}}

\begin{document}

{\bf\large \noindent The NGC 1023 Galaxy Group: An Anti-Hubble
Flow?} \vspace{0.5cm}

{ A. D. Chernin$^{1,2}$, V. P. Dolgachev$^{1}$, L. M.
Domozhilova$^{1}$ }

\vspace{1cm}

{\it $^{1}$Sternberg Astronomical Institute, Moscow University,
Moscow, Russia,

 $^2$Tuorla Observatory, University of Turku, Finland, 21 500}

\vspace{1cm}

(Astronomy Reports, 2010, Vol. 54, No. 10, pp. 902–907; in
Russian: Astronomicheski Zhurnal, 2010, Vol. 87, No. 10, pp.
979–985)

\vspace{1cm}

{\bf Abstract}

\vspace{1cm}

We discuss recently published data indicating that the nearby
galaxy group NGC 1023 includes an inner virialized
quasi-stationary component and an outer component comprising a
flow of dwarf galaxies falling toward the center of the system.
The inner component is similar to the Local Group of galaxies, but
the Local Group is surrounded by a receding set of dwarf galaxies
forming the very local Hubble flow, rather than a system of
approaching dwarfs. This clear difference in the structures of
these two systems, which are very similar in other respects, may
be associated with the dark energy in which they are both
imbedded. Self-gravity dominates in the Local Group, while the
anti-gravity produced by the cosmic dark-energy background
dominates in the surrounding Hubble flow. In contrast,
self-gravity likewise dominates throughout the NGC 1023 Group,
both in its central component and in the surrounding “anti-Hubble”
flow. The NGC 1023 group as a whole is apparently in an ongoing
state of formation and virialization. We may expect that there
exists a receding flow similar to the local Hubble flow at
distances of 1.4–3 Mpc from the center of the group, where
anti-gravity should become stronger than the gravity of the
system.

\vspace{1cm}

\section{Introduction}

It has been known since the time of Hubble’s observations [1] that
the nearby Universe to distances of  about 10-15 Mpc is populated
by comparatively small groups of galaxies similar to the Local
Group. Essentially all major galaxies and most known dwarf
galaxies in this volume are collected in such groups [2]. About
two dozen dwarf galaxies are observed around the Local Group,
which are moving away from the center of the group. This expanding
component of the system represents the nearby local Hubble flow of
receding galaxies (see the new study [3] and references therein).
For brevity, we will call such a system, with a central,
quasi-stationary group surrounded by an expanding flow, a “Hubble
cell”. In addition to the local cell [3–9], the Hubble cells
around the giant galaxies M81 and Cen A have also been studied
[10, 11]. Hubble cells can be considered the main structural unit
of the nearby Universe. A theoretical model for a Hubble cell was
proposed and developed in [4–9, 12–14]. The main new aspect of
this model is that it takes into account the cosmic dark-energy
background. In accordance with the standard LambdaCDM cosmology,
the model assumes that a group of galaxies together with the
surrounding flow of the Hubble recession is imbedded in a uniform
dark energy distribution that is constant in time. Dark energy
creates anti-gravity. The gravity due to baryonic and dark matter
in the group dominates within the volume of the group, while the
anti-gravity of the dark energy dominates in the area of the
Hubble flow; anti-gravity becomes stronger than the gravity of the
group at the distances of 1–3 Mpc from the group center. The
boundary between the gravitationally bound, quasi-stationary group
and the expanding Hubble flow corresponds to the "zero-gravity
surface", where the forces of gravity and anti-gravity cancel.

It may be seemed at the first glance that the kinematic structure
of the NGC 1023 Group represents a counter-example to this model
for a Hubble cell. According to Trentham and Tully [15], there is
a gravitationally bound (and virialized) system of dwarf galaxies
in the central region of this group, surrounding the giant galaxy
NGC 1023, which is the most massive galaxy in the group. A flow of
dwarf galaxies is observed outside the central system. This
two-component structure is reminiscent of the Local Group, with
its flow of receding galaxies. However, Trentham and Tully [15]
claim that, in the case of NGC 1023, the flow represents an infall
toward the center of the group, rather than an outflow away from
it.

In this paper, we discuss the kinematic and dynamical structures
of the NGC 1023 Group based on the data [15–19], and evaluate
their agreement with our Hubble-cell model. In Section 2, a brief
description of a typical Hubble cell, such as the Local Group
together with its surrounding local Hubble flow, is given. Section
3 presents data on the NGC 1023 Group and considers the kinematics
of its outer component. In Section 4, we consider the dynamical
background against which the “anti- Hubble” flow of the group
develops. Our results are discussed in Section 5. 2.

\vspace{1cm}

\section{ The Local Hubble Cell}

The Local Group of galaxies contains two giant galaxies — the
Milky Way and M31. The group also includes the Magellanic clouds,
M33, and approximately 50 dwarf galaxies. The Local Group is a
quasi-stationary, gravitationally bound system immersed in a
potential well created primarily by the gravitational force of the
dark matter collected in the extended massive halos of the two
giant galaxies in the group. The total mass of the group (baryonic
and dark matter) is estimated to be $M_{LG} \simeq (1-5)\times
10^{12} M_{\odot}$ [3, 20, 21, 22]. The diameter of the Local
Group is approximately 2 Mpc, while the distance between the
centers of the Milky Way and M31 is about 0.7 Mpc. These two
galaxies (together with the families of dwarf galaxies populating
their individual dark halos) are approaching each other with the
velocity that is now near 120 km/s. The onset of the local Hubble
flow is just outside the Local Group, at distances of R > 1.4-1.6
Mpc from its center [3]. The flow is formed by 22 dwarf galaxies
located at distances of up to 3 Mpc. All the dwarf galaxies of
this flow are receding from the group, as a rule, with velocities
(relative to the group barycenter) that increase with distance
from the center of the group.

The local Hubble cell — the Local Group together with the local
Hubble flow — serves as a typical example of a Hubble cell. In the
velocity–distance diagram presented in [3], 58 galaxies of the
cell occupy two well defined regions. The first of these,
corresponding to distances to 1.3–1.5 Mpc, is occupied by galaxies
of the Local Group. Galaxies in this region have (radial)
velocities that are both positive and negative, in the interval
from -150 to +170 km/s, with the mean velocity being close to
zero. The mean (radial) velocity dispersion of the galaxies in the
system is 72 km/s. The second component of the system is the local
Hubble flow, where there is no negative velocities, and the
velocities of galaxies have values from 60 km/s near the distances
of 1.6 Mpc to 250 km/s at the distance of 3 Mpc.

Two other Hubble cells studied in detail are around the М81 galaxy
and the Cen A galaxy [10,11]; they are similar to each other and
to the Local Cell. Their central groups are similar: in each of
them there is a dominant galaxy or pair of massive galaxies. The
local Hubble flows around the groups are even more similar. The
flows are formed by dwarf galaxies whose total mass in each case
is much less than the mass of the central group. The flows display
a high degree of regularity, and a nearly linear velocity–distance
dependence, with the median local Hubble factor which is in a
narrow range: 57 km/s/Mpc < $H_{med}$ < 62 km/s/Mpc.

\vspace{1cm}

\section{The NGC 1023 Group: two components}

The NGC 1023 Group, first identified in [23], is sometimes
considered a “classic example” [24, 25] of a small system of
galaxies. It is located in the direction opposite to the Virgo
Cluster, and the group is compact and well isolated in space [16].
Tully [25] identified 14 members of the group, for five of which
distances were determined with the use of the Tully–Fisher
relation. Distances to 11 galaxy in the group were derived in [16]
based on data obtained on the 6-m telescope of the Special
Astrophysical Observatory and the Hubble Space Telescope (nine
using the brightest-stars method and two based on the tip of the
red-giant branch). In their recent paper “Dwarf Galaxies in the
NGC 1023 Group,” Trentham and Tully [15] present data on 70
galaxies in the group, 65 of which were studied using the MegaCam
detector on the 3.6-m Canada-France-Hawaii Telescope (CFHT). The
main galaxy in the group is the lenticular giant NGC 1023, whose
distance is estimated to be from 9.86 Mpc [16] to 11.4 Mpc [25].
Its velocity of recession from the barycenter of the Local Group
is 828 km/s [18], and its heliocentric velocity is 637 km/s [15].
Morphological types, Galactic coordinates and angular distances
from the main galaxy are given for all the galaxies in [15];
(heliocentric) velocities are available for 25 galaxies, with all
velocities lying within 1000 km/s.

We use the data from [15] to construct a heliocentric
velocity–angular distance diagram for 27 galaxies in the NGC 1023
Group (Fig.1) which reproduces (in a modified form, see below)
Fig.8 of [15]. Note that, as in [15, Fig. 8], our Fig.1 is not a
Hubble diagram. A standard Hubble diagram is constructed using
radial velocities and radial distances for the galaxies taken
relative to the barycenter of the system. In contrast to this, the
velocities in Fig.1 are heliocentric; the vertical axis plots the
difference between the velocity of each galaxy and the velocity of
the main galaxy in the group (637 km/s). A positive velocity in
Fig.1 indicates motion toward (or away from) the center of the
group, if the given galaxy is located closer to us (or further
from us) than the main galaxy in the group. Correspondingly, a
negative velocity indicates motion toward (or away from) the
center of the given, if the galaxy is located further from us (or
closer to us) than the main galaxy. The horizontal axis plots the
distance of each galaxy from the main galaxy projected onto the
plane of sky. In accordance with [15, 16], the distance of the
main galaxy NGC 1023 from the Sun is taken to be 10 Mpc.

The two-component structure of the group is clearly seen in this
diagram. The central component is comprised of 20 galaxies,
including NGC 1023. Seventeen of these galaxies were taken from
[15] and three from the studies [18, 19], which report the
discovery of 12 new galaxies of the NGC 1023 group. Heliocentric
velocities for three of the galaxies were measured on the 100-m
Effelsberg radio telescope. The dwarf galaxies of the central
component are concentrated around the main galaxy NGC 1023, and
have both positive and negative velocities (or more precisely,
velocity differences), from -200 to +300 km/s. The object DDO 22 =
UGC 2014, which was considered a member of the central component
in [15], is marked by a special symbol in our diagram (a circle
with a bar). According to [16], its distance from the Sun is
relatively large, $17 \pm 2$ Mpc, making it unlikely that it is
indeed a member of the group. By analogy with the Local Group (see
Section 2), we expect that the central component of the NGC 1023
group is gravitationally bound and quasistationary. As was
proposed in [15], the dwarfs of the central component move along
finite orbits within the massive and extended dark halo of the
main galaxy. With a velocity dispersion for the dwarfs of 140 km/s
and a mean harmonic radius of 0.3 Mpc, a standard virial estimate
yields for the mass of the central component $M = (6 \pm 3)\times
10^{12} M_{\odot}$ [15]. The luminosity of all 40 galaxies within
a volume with a radius of 0.3 Mpc (not only the 16 included in
[15, Fig. 8]) is $2 \times 10^{10} L_{\odot}$ [15], so that the
mass-tolight ratio is approximately 300 in solar units. The dwarfs
of the central subsystem considered in [15] are predominantly
elliptical galaxies; the three dwarfs from [18, 19] are irregular
galaxies.

The outer component in Fig.1 contains 10 galaxies with velocities
from -100 to 0 km/s. Their distances from the central galaxy in
Fig.1 are 0.4–1.4 Mpc. As was proposed in [15], these galaxies
form a flow directed toward the center of the group, and their
motion represents a first infall of these objects in the
gravitational field of the central component. The mean flow
direction is essentially parallel to the horizontal axis, and
corresponds to a velocity of roughly -60 km/s. The galaxies in the
flow have primarily late morphological types. The galaxy DDO 19 =
UGC 1865 in the area of the flow is shown in our diagram by a
circle with a bar; according to [16], its distance, $30 \pm 3.6$
Mpc, is too large to make it likely that it is a member of the
group. The negative velocities (velocity differences) of the outer
component indicate that these galaxies are falling toward the
center of the group—but only if their distances from the Sun are
greater than the distance to the main galaxy. This was explicitly
assumed in [15], however no radial-distance data are presented in
[15].

The question of the radial distances of the galaxies in the flow
is of critical importance for our understanding of the real
kinematics of the outer component of the NGC 1023 Group. The
measurements of [16, 17] provide distances for six galaxies of the
flow (including UGC 1865). The distance to DDO 25 = UGC 2023 is
$7.7 \pm 0.9$ Mpc. Three other galaxies — NGC 959, NGC 925, and
NGC 891 — have distances of $9.3 \pm 1$, $8.9 \pm 0.25$, and $9.82
\pm 0.25$ Mpc, respectively. If these four galaxies are indeed
closer to the Sun than the main galaxy (recall that the distance
to NGC 1023 has been estimated to be from 9.86 [16] to 11.4 Mpc
[25]), they are not moving toward the center of the group, but
instead away from the center. The fifth of the flow galaxies
indicated above — NGC 949 — is located at a distance of $14.5 \pm
1.7$ Mpc, which probably exceeds the distance to NGC 1023. In this
case, its negative velocity and relatively large radial distance
indicates motion toward the center of the group.

How reliable are these six distances? The distances for two of the
galaxies — NGC 891 and NGC 925 — are based on the tips of their
red-giant branches [16, 17]; in this case, the accuracy of these
measurement are good (appreciably better than 10\%), and these
distances should be quite reliable. The same method was used to
measure the distance to the main galaxy of the group with the same
accuracy and reliability ($9.86 \pm 0.25$ Mpc) [16, 17]. The
distances to the other galaxies of the flow were measured using
the brightest-supergiant method. In this case, there are
appreciably systematic uncertainties: the distances to the dwarf
galaxies depend substantially on whether they contain bursts of
star formation [27]. This circumstances remains an essentially
unverifiable source of uncertainty. It follows that there are
actually only two reliable distances for the flow component of the
NGC 1023 Group, with both of these (see above) being close enough
to the distance of the main galaxy, so that, even with the high
accuracy of these measurements, the three distances are the same
within the errors. Thus, we are not able to elucidate with
certainty where these two galaxies are located closer or further
than the main galaxy of the group. For this reason, the question
of the direction of their motion and the direction of the entire
flow remains open.

\section{The NGC 1023 Group: dynamical background}

In an attempt to clarify  the dynamical situation in the group, we
will consider a simple model for the system which enables us to
estimate the gravitational and anti-gravitational forces in the
volume of the group. Following the general approach adopted to
study nearby galaxy groups (see Sections 1 and 2), we assume that
the NGC 1023 group (like the Local Group) is imbedded in a
distribution of dark energy, whose density is uniform in space and
constant in time. We also assume that the gravitational force in
the group is due mainly to the mass of dark matter in the
spherical halo of its main galaxy, and that the dwarf galaxies of
the group can be treated approximately as test particles. The
dynamics of the group can be described using Newtonian mechanics;
the relativistic properties of dark energy and the anti-gravity it
creates can adequately be formulated using the language of
classical forces and potentials. In this model, two forces act on
each of the test particles in the group: the gravity due to the
central mass, and the anti-gravity due to the dark-energy
background. In a frame of the barycenter of the mass, the
gravitational force (per unit particle mass) is nearly exactly
central, since the dark halo of the main galaxy can be taken to be
approximately spherical; this force is given by the law of
Newtonian gravitation,

\be F_N = -GM/R^2, \ee

where $G$ is the gravitational constant, $M$ the central mass, and
$R$ the distance of the particle from the center of mass. In this
frame, the anti-gravitational force (which is exactly central) is
determined by the density of dark-energy$\rho_V$ , and is given
(also per unit mass) by the “law of Einstein anti-gravitation:”

\be F_E = G2\rho_V (4\pi/3)R^3/R^2 = (8\pi/3) G \rho_V R.\ee

As we can see from (1) and (2), the gravitational and
anti-gravitational forces dominate at small and large distances,
correspondingly. The forces are equal on the absolute value at the
distance

\be R = R_V = (\frac{3M}{8\pi \rho_V})^{1/3}, \ee  where the total
acceleration of the particles vanishes. Gravity dominates when $R
< R_V$ and anti-gravity when $R > R_V$. Adopting that the
dark-energy density has the value $\rho_V = 0.72 \times 10^{-29}$
g/cm$^3$, as measured in global cosmological observations [28].
The zero-gravity radius then

\be R_V \simeq 1 \times (M/10^{12} M_{\odot})^{1/3} Mpc. \ee

Substituting the mean virial mass for the central component of the
group, $M = 6\times 10^{12} M_{\odot}$, we find that the critical
value $R_V$ for this system is 1.8 Mpc. Given the uncertainties
(see above), the estimated mass is probably in the range
$(3-9)\times 10^{12} M_{\odot}$, the zero-gravity radius is then
$R_V = 1.4-2.1$ Mpc.

It follows from these estimates and the data on the distances
(angular sizes) of the group components (see Fig.1) that the mean
value of the critical radius, $R_V = 1.6$ Mpc, exceeds the
distance to the galaxy flow; it means that both components of the
group prove to be contained within the zero-gravity sphere. This
condition is satisfied with a substantial margin for the highest
mass estimate for the system. But even for the lowest mass
estimate, the observed projected distance to the most distant part
of the galaxy flow does not exceed $R_V$. Thus, the outer
component of the system is within the region where gravity
dominates. This is the principle difference of the NGC 1023 Group
from the Local Hubble Cell: in the NGC 1023 Group, both the
central component and the flow are located inside the zero-gravity
sphere, while, in the Local Hubble Cell, the central component
(Local Group) and the flow are located on different sides of this
spherical surface. The fact that gravity dominates in the outer
component of the NGC 1023 Group does not mean that anti-gravity
plays no role in the dynamics of this component. The dark-energy
background reduces the effective gravitating mass, $M_{eff} = M -
(8\pi/3)\rho_V R^3$, acting on the particles of the outer
component. If, for example, we adopt for the mean virial mass $M =
6\times 10^{12} M_{\odot}$, the effective mass for a galaxy that
is most distant (in terms of angular distance) from the center (a
projected distance of $R = 1.4$ Mpc) is roughly half the virial
mass $M$. If the virial mass corresponds to its lowest estimate,
$M = 3\times 10^12 M_{\odot}$ (see above), the effective mass is
close to zero at the same distance.

It is most likely that the presence of the outer component in the
NGC 1023 Group is a manifestation of ongoing formation of the
group. This process proceeds in a non-linear regime, and is
developing in the region where gravity dominates over
anti-gravity. The anti-gravity is able to slow this process down,
but not to stop it.

\section{Conclusion}

In the standard $\Lambda$CDM cosmological model, dark energy is
described by Einstein cosmological constant. If this is the case,
then dark energy should be present everywhere in space, and have a
density that is constant in space and time. Applied to the nearby
Universe, this means that the Local Hubble Cell and other similar
systems of the spatial scales of 1–3 Mpc are imbedded in the
uniform cosmological dark-energy background, and should be subject
to its anti-gravitational force.

In the dynamics of the galaxy group, the antigravity of the dark
energy is able to compete with the gravity due to the baryonic and
dark matter of the galaxies. A gravitationally bound system is
possible only if gravity is stronger than anti-gravity within its
volume. The zero-gravity radius $R_V$ is a critical quantity: the
volume of a quasi-stationary gravitationally bound virialized
system cannot extend beyond a sphere with this critical radius. As
has been elucidated [12–14, 16], the condition $R < R_V$ is
satisfied in the Local Group of galaxies, while the surrounding
Local Hubble flow is located in the region where antigravity
dominates: the distance to the flow from the barycenter of the
group exceeds the critical value ($R > R_V$). Where anti-gravity
dominates, it tries to make particles recede from the central mass
(and from each other) with acceleration. With time, a close to
linear dependence of the velocity on distance is established in
the flow (in a frame that is fixed to the group barycenter).

The NGC 1023 Group is very similar to the Local Group: its mass
and size are close to those of the Local Group. However, details
of the internal structure of the groups differ significantly: the
Local Group is in a state close to virial equilibrium throughout
all its volume, while only the central component of the NGC 1023
Group is close to this state. Virial equilibrium does not extend
to the outer component of the NGC 1023 group, where about ten
galaxies are contained. As was proposed in [15], the outer
component represents a flow of galaxies directed toward the group
center. However, we have shown (see Section 3) that this is not
supported by all the available (rather sparse) data. However,
whatever the direction of the galaxy velocities for this
component, it is clearly not an “anti-Hubble” flow, since the
outer component is located inside the sphere of the gravity
domination (see Section 4). This last circumstance can serve as
independent, although indirect, support for the suggestion of [15]
that the outer component is falling into the center: since gravity
dominates throughout the volume of the group, this direction of
motion seems preferred.

Summarizing, we emphasize that the NGC 1023 group is in no way a
counter-example to our model of the Hubble cell. The structure and
dynamics of the group in does not contradict the model. The
two-component NGC 1023 Group most likely represents the central,
gravitationally bound region of a more or less standard Hubble
cell. If this is the case, we may expect that there should be a
flow of receding dwarf galaxies in the vicinity of the group. This
receding flow could be observed at distances (from the center of
the group) exceeding the zero-gravity radius ($R > R_V \simeq
1.4-2$ Mpc), in the area where the anti-gravity due to the cosmic
dark-energy background dominates. This hypothesis can be verified
by searching for further galaxies in the vicinity of the group and
accurately measuring their velocities and distances. This would be
an interesting observational problem, directly related to the
local dynamical effects of dark energy.

We thank O.K. Sil’chenko, J. Vennik, I.D. Karachentsev,
P.Teerikorpi and N.A. Tikhonov for useful advice and critical
comments.

\section*{REFERENCES}

 1. E. P. Hubble, The Realm of Nebulae (New Haven, 1936).

2. I. D. Karachentsev, Astron. J 129, 178 (2005).

3. I. D. Karachentsev, O. G. Kashibadze, D. I. Makarov, and R. B.
Tully, Mon. N ot. R. Astron. Soc. 393, 1265 (2009).

4. A. D. Chernin, I. D. Karachentsev, M. J. Valtonen, et al.,
Astron. Astrophys. 415, 19 (2004).

5. P. Teerikorpi, A. D. Chernin, and Yu. V. Baryshev, Astron.
Astrophys. 440, 791 (2005).

6. P. Teerikorpi, A. D. Chernin, and Yu. V. Baryshev, Astron.
Astrophys. 483, 383 (2006).

7. G. G. Byrd, A. D. Chernin, and M. J. Valtonen, Cosmology:
Foundations and Frontiers (URRS, Moscow, 2007).

8. A. D. Chernin, I. D. Karachentsev, M. J. Valtonen, et al.,
Astron. Astrophys. 467, 933 (2007).

9. A. D. Chernin, P. Teerikorpi, and Yu.V. Baryshev, Adv. Space
Res. 31, 459 (2003); arXiv: astro-ph/0012021 (2000).

10. A. D. Chernin, I. D. Karachentsev, D. I. Makarov, et al.,
Astron. Astrophys. Trans. 26, 275 (2007).

11. A. D. Chernin, I. D. Karachentsev, O. G. Kashibadze, et al.,
Astrofizika 50, 405 (2007).

12. A. D. Chernin, Usp. Fiz. Nauk 171, 1153 (2001) [Phys. Usp. 44,
1099 (2001)].

13. P. Teerikorpi,A. D. Chernin, I. D. Karachentsev, et al.,
Astron. Astrophys. 483, 383 (2008).

14. A. D. Chernin, Usp. Fiz. Nauk 178, 267 (2008) [Phys. Usp. 51,
253 (2008)].

15. N . Trentham and R. B. Tully, Mon. N ot. R. Astron. Soc. 398,
722 (2009).

16. N. A. Tikhonov and O. A. Galazutdinova, Astrofizika 45, 311
(2002).

17. N. A. Tikhonov and O. A. Galazutdinova, Astrofizika 48, 261
(2005).

18. I .D. Karachentsev, V. E. Karachentseva, and W. K. Huchtmeier,
Pis’ma Astron. Zh. 33, 577 (2007) [Astron. Lett. 33, 512 (2007)].

19. W. K. Huchtmeier, I. D. Karachentsev, and V. E. Karachentseva,
Astron. Astrophys. 506, 677 (2009).

20. S. van den Bergh, Astron. J. 124, 782 (2002).

21. S. van den Bergh, Astrophys. J. Lett. 559, L113 (2001).

22. A. D. Chernin, P. Teerikorpi, M. J. Valtonen, et al., Astron.
Astrophys. 507, 1271 (2009).

23. M. L. Humason, N. U. Mayall, and A. R. Sandage, Astron. J. 61,
97 (1956).

24. J. Materne, Astron. Astrophys. 33, 451 (1974).

25. R. B. Tully, Astrophys. J. 237, 390 (1980).

26. J. L. Tonry et al., Astrophys. J. 546, 681 (2000).

27. N. A. Tikhonov, private commun. (2010).

28. D. N . Spergel, R. Bean, O. Dore. , et al., Astrophys. J.
Suppl. Ser. 170, 377 (2007).

\section*{Figure caption}

Velocity–distance diagram for 30 galaxies of the NGC 1023 Group
based on the data of [15-18]. This is not a standard Hubble
diagram: the velocities and distances have non-standard meanings.
The velocity of each galaxy is the difference between its radial
heliocentric velocity and the radial velocity of the main galaxy
in the group (637 km/s). The distances shown are the projection of
the distance of each galaxy from the main galaxy onto the plane of
the sky; the line-of-sight distance from the Sun to the main
galaxy is taken to be 10 Mpc. The solid lines show the boundaries
of the central quasi-stationary component of the group (distances
$<0.4$ Mpc) and the outer component (from $0.5$ to $1.4$ Mpc),
which is considered in [15] to be a flow directed toward the
center of the group. The open circles denote galaxies from [15]
for which CFHT MegaCam data are available, the pluses denote
galaxies from [15] for which such data were not obtained, and the
filled circles galaxies from [18]. The two galaxies shown as
circles with bars most likely lie outside the group.

\end{document}